\documentstyle[12pt]{article}
\begin{document}
\bibliographystyle{plain}
\renewcommand{\theequation}{\arabic{section}.\arabic{equation}}
\renewcommand{\arraystretch}{1.4}
\newcommand{\beq}{\begin{equation}}
\newcommand{\eeq}{\end{equation}}

\newcommand{\bea}{\begin{eqnarray}}
\newcommand{\eea}{\end{eqnarray}}
\newcommand{\tr}{\mbox{\rm str}}

\newcommand{\bean}{\begin{eqnarray*}}
\newcommand{\eean}{\end{eqnarray*}}
\def\cg{{\cal G}}
\def\ck{{\cal K}}
\def\cl{{\cal L}}
\def\n2{{$N=2$}}
\newcommand{\reff} [1]
	{(\ref{#1})}

\renewcommand{\thefootnote}{\alph{footnote})}
\newcommand{\un}{\underline}
\oddsidemargin=0.25cm\evensidemargin=0.25cm
\thispagestyle{empty}
\begin{flushright}ENSLAPP-L-559\\
hep-th/9512220\\
December 1995
\end{flushright}
\vskip 1.0truecm
\begin{center}{\bf\Large
Gauge invariant formulation of \n2 Toda\\
and KdV systems in extended superspace}
\end{center}
 \vskip 1.0truecm
\centerline{\bf F. Delduc, M. Magro}
\vskip 1.0truecm
\begin{center} \it Laboratoire de Physique Th\'eorique ENSLAPP
\begin{footnote}{URA 14-36 du CNRS, associ\'ee  \`a l'ENS de Lyon et au LAPP\\
\hspace*{.6cm}Groupe de Lyon: ENS Lyon, 46 All\'ee d'Italie, 69364 Lyon, France
}\end{footnote}\end{center}
\vskip 2.0truecm  \nopagebreak

\begin{abstract}
We give a gauge invariant formulation of $N=2$ supersymmetric
abelian Toda field equations in \n2 superspace. Superconformal invariance is
studied. The conserved currents are shown to be associated with
Drinfeld-Sokolov type gauges. The extension to non-abelian \n2 Toda equations
is discussed.
Very similar methods are then applied to a matrix formulation in \n2
superspace of one of the \n2 KdV hierarchies.
\end{abstract}
\setcounter{page}0
\newpage

\section*{Introduction}

The \n2 supersymmetric Liouville equation together with its Lax representation
in superspace was first given in \cite{ivan2}. The generalization to
the abelian Toda equations   have been derived and studied in
\cite{EH}. These models are associated with the principal embedding of the
superalgebra $sl(2\vert 1)$ inside $sl(n+1\vert n)$. Here we construct a set of
gauge invariant equations in \n2 superspace which, in a particular gauge,
reduce to the  \n2 supersymmetric abelian Toda equations.
This is in the spirit of the works \cite{FOR}
and \cite{DRS}, where such a gauge invariant formulation was constructed for
bosonic Toda models in ordinary space, and for supersymmetric Toda models
in $N=1$ superspace. The $N=1$ superspace approach leads to a natural
interpertation for the use of superalgebras admitting a principal $osp(1\vert
2)$ embedding \cite{LS}. The \n2 abelian Toda equations correspond to the cases
when this principal $osp(1\vert 2)$ embedding may be extended to an $sl(2\vert
1)$ embedding. The consistency of the gauge invariant equations is ensured by
an interesting interplay between the geometry of the extended superspace and
the structure of the
$sl(2\vert 1)$ superalgebra. Such a gauge invariant formulation allows for an
easy discussion of the conserved currents of the Toda equations.
However, it should be stressed that the formulation we give in this article
is restricted to the discussion of field equations. We do not have an
explicitly supersymmetric action. A hamiltonian formulation in extended
superspace is presently out of reach\footnote{It is to be noted that  a
Hamiltonian reduction approach to \n2 $\cal W$
algebras was developped in \cite{ivan2}.}. Another difference with previous
works is that in the \n2 superspace, it seems hard to interpret the gauge
invariant field equations as a gauged WZNW model.

The paper is organized as follows. In the first section we recall some basic
facts about the superalgebra $sl(n+1\vert n)$ and its principal $sl(2\vert 1)$
embedding. In section 2, we write down the gauge invariant field equations and
establish the relation with \n2 Toda equations. Then we discuss the conserved
currents of \n2 Toda equations and their
superconformal transformations. Finally we apply this formalism to
non-abelian \n2 Toda equations. In section 3, we construct in \n2 superspace
a matrix Lax formulation of an \n2 supersymmetric KdV hierarchy.

\section{The superalgebra $sl(n+1\vert n)$ and its principal  $sl(2\vert 1)$
subalgebra }
\setcounter{equation}{0}
This section is devoted to a short introduction to the super-algebras
$sl(n+1\vert n)$. We shall choose a basis which allows for an easy description
of the principal $sl(2\vert 1)$ subalgebra.\\

We consider the set of $(2n+1)\times(2n+1)$ real matrices. A convenient basis
is given by the matrices $E_{i,j}$ such that
$(E_{i,j})_{kl}=\delta_{ik}\delta_{jl}$. We define the supertrace of a matrix
$M$ by the alternate sum
\beq \tr M=\sum_{i=1}^{2n+1}(-1)^{i+1}M_{ii}. \eeq
The supercommutator of two matrices $M$ and $N$ is
\beq
[M,N\}_{ij}=\sum_{k=1}^{2n+1}(M_{ik}N_{kj}-(-1)^{(i+k)(k+j)}N_{ik}M_{kj}).
\eeq
One then easily checks that $\tr[M,N\}=0$. The superalgebra
$\cg=sl(n+1\vert n)$ is the set of matrices with zero supertrace. A Cartan
subalgebra of $\cg$ is generated by the diagonal matrices
\beq H_i=E_{i,i}+E_{i+1,i+1}, \,\,\, 1\leq i\leq 2n.
\eeq
The superalgebra $\cg$ is $\mbox{Z}_2$ graded, $\cg=\cg_{\bar 0}
+\cg_{\bar 1}$. The $\mbox{Z}_2$-grading of a matrix $E_{i,j}$ is
$i+j$ modulo $2$. We shall denote by a hat the superalgebra automorphism which
reverses the sign of odd elements,
\beq (\hat M)_{ij}=(-1)^{i+j}M_{ij}  .\eeq
The matrices $E_{i,i+1}$ just above the diagonal are associated with a complete
set of fermionic simple roots. We now define the principal $sl(2\vert 1)$
embedding. The Cartan subalgebra is spanned by
\beq H=\sum_{i=1}^n{n-i+1\over 2} H_{2i-1},\,\,\,{\bar H}=
-\sum_{i=1}^n{i\over 2}H_{2i} .\eeq
The matrices $\mu_+$, $\bar\mu_+$ associated with positive simple fermionic
roots are
\beq \mu_+=\sum_{i=1}^nE_{2i-1,2i},\,\,\,\,\bar\mu_+=\sum_{i=1}^n E_{2i,2i+1},
\eeq
and the matrices $\mu_-$, $\bar\mu_-$ associated with negative simple fermionic
roots are
\beq \mu_-=\sum_{i=1}^n(n-i+1)E_{2i,2i-1},\,\,\,\,\bar\mu_-=-\sum_{i=1}^n
iE_{2i+1,2i}.
\eeq
The non-zero super-commutators are
\bea &
[\bar H, \mu_\pm]=\pm {1\over 2}\mu_\pm,\,\,\,
[H,\bar\mu_\pm]=\pm{1\over 2}\bar\mu_\pm  &\\
&\{\mu_+,\mu_-\}=2H,\,\,\,\{\bar\mu_+,\bar\mu_-\}=2\bar H, &\\
&\{\mu_+,\bar\mu_+\}=E_{++},\,\,\,
\{\mu_-,\bar\mu_-\}=E_{--},  \label{eq15}&\eea
 where $E_{++}$ and $E_{--}$ are associated with the bosonic roots of
$sl(2\vert 1)$. This algebra contains the principal $osp(2\vert 1)$ subalgebra
of $\cal G$.
The Cartan generator is $e_0=H+\bar H$, the operators associated with positive
and negative fermionic simple roots are $f_+=\mu_++\bar\mu_+$,
$f_-=\mu_-+\bar\mu_-$. The operator $\mbox{ad}e_0$ is diagonalizable. The
eigenvalues are ${i\over 2}$, $-n\leq i\leq n$. We denote the corresponding
eigenspaces by $\cg_{i\over 2}$. This defines a half-integer grading of $\cal
G$, $\cg=\oplus_{i=-n}^n\cg_{i\over 2}$. In the basis that we have chosen,
positive (negative) eigenvalues correspond to upper (lower) triangular matrices
and $\cg_0$ is the Cartan subalgebra. Even elements have integer gradation, and
odd elements have half-integer gradation. In particular, $\mu_+$ and
$\bar\mu_+$ ($\mu_-$ and $\bar\mu_-$) belong to the eigenspace $\cg_{1\over 2}$
($\cg_{-{1\over 2}}$).

Finally, in the following we always use matrices taking values in some
Grassmann algebra $\cg r=\cg r_{\bar 0}\oplus\cg r_{\bar 1}$. These matrices
will be called even if they belong to
${\cal E}=\cg_{\bar 0}\otimes\cg r_{\bar 0}\oplus\cg_{\bar 1}\otimes\cg r_{\bar
1}$
and odd if they belong to
${\cal O}=\cg_{\bar 0}\otimes\cg r_{\bar 1}\oplus\cg_{\bar 1}\otimes\cg r_{\bar
0}$.
We then define the supercommutator of two matrices $A$ and $B$ belonging to
${\cal E}$ or ${\cal O}$ by
\beq [A,B\}=\sum_{i,j,k,l}A_{ij}B_{kl}[E_{i,j},E_{k,l}\}\eeq
In the various cases, this takes the matrix product forms listed in the table
below.
\begin{center}
\begin{tabular}{|c|c|c|}\hline
$[A,B\}$ & $A\in {\cal E}$ & $A\in {\cal O}$ \\ \hline
$B\in {\cal E}$ & $AB-BA$ & $AB-\hat BA$ \\ \hline
$B\in {\cal O}$ & $AB-B\hat A$ & $AB+\hat B\hat A$\\ \hline
\end{tabular}
\end{center}
\section{Gauge invariant formulation of the Toda field equations}
\setcounter{equation}{0}
\label{eq100}
The method used in this section is an extension of that developped in
\cite{FOR}. However, our method only applies to the field equations, and we
do not have an explicitly supersymmetric action or Hamiltonian formulation.

\subsection{Zero curvature equations and constraints}

We denote the supergroup corresponding to the superalgebra
$\cg =sl(n+1\vert n)$ by $G=SL(n+1\vert n)$. We use the grading described in
the last section, and separate $\cg$ as
$\cg=\cg_{<0}\oplus\cg_0\oplus\cg_{>0}$.
$\cg_{>0}$ contains all upper triangular matrices, $\cg_{<0}$ all lower
triangular matrices. $\cg_0$ is the Cartan subalgebra spanned by diagonal
matrices. The supergroups corresponding respectively to $\cg_{<0}$,
$\cg_0$, $\cg_{>0}$ will be denoted by $G_{<0}$, $G_0$, $G_{>0}$. The elements
of $G_{>0}$ ($G_{<0}$) are upper (lower) triangular even matrices
with ones on the diagonal.
The coordinates of the \n2 superspace are the light-cone coordinates $(x^{++},
x^{--})$ together with the Grassmann coordinates
$(\theta^+,\bar\theta^+,\theta^-,\bar\theta^-)$. Notice that we use an \n2
supersymmetry algebra with $SO(1,1)$ automorphism group. The covariant spinor
derivatives are
\beq D_+={\partial\over\partial\theta^+}+\bar\theta^+\partial_{++},\,\,
\bar D_+={\partial\over\partial\bar\theta^+}+\theta^+\partial_{++},\,\,
\partial_{++}={\partial\over\partial x^{++}},
\eeq
and $D_-$, $\bar D_-$ are similarly defined. These derivatives satisfy the
algebra
\beq
D_+^2=\bar D_+^2 =0,\,\, \{D_+,\bar D_+\}=2\partial_{++},\,\,\{\tilde
D_+,\tilde D_-\}=0.
\eeq
where we use a tilde as a generic notation for unbarred or barred objects,
$\tilde D_+=D _+$ or $\bar D_+$.

We now introduce the superfields that we need for the gauge invariant
formulation of Toda equations. The first is $g(x,\theta,\bar\theta)$ which
takes values in the supergroup $G$.
Then there are spinor gauge superfields $A_\pm(x,\theta,\bar\theta)$, $\bar
A_\pm(x,\theta,\bar\theta)$.
$A_+$ and $\bar A_+$ are upper triangular odd matrices, $A_-$ and $\bar A_-$
are lower triangular odd matrices. The gauge transformations
$R(x,\theta,\bar\theta)\in G_{>0}$ and $L(x,\theta,\bar\theta)\in G_{<0}$
act on the superfield $g(x,\theta,\bar\theta)$ by left and right translations
\beq
g\longrightarrow LgR,\eeq
and on the gauge superfields by
\bea
\tilde A_+&\longrightarrow& \hat R^{-1}\tilde D_+ R+\hat R^{-1}\tilde A_+ R,\\
\tilde A_-&\longrightarrow& \tilde D_- L L^{-1}+\hat L\tilde A_- L^{-1}.
\eea
At that point, the connexions $\tilde A_+$ and $\tilde A_-$ transform under
different gauge groups. However, the field $g$ acts as a ``bridge'' between the
two gauge groups and may be used to construct connexions transforming under the
same gauge group,
\beq\tilde B_+=-\hat g\tilde A_+g^{-1}+\tilde D_+ gg^{-1},\,\,\, \tilde B_-=
\tilde A_-,\eeq
The connexion $\tilde B_\pm$ only transform under the left gauge group
\beq \tilde B_\pm\longrightarrow (\tilde D_\pm L)L^{-1}+\hat L\tilde B_\pm
L^{-1} \eeq
One could equivalently use a set of connexions transforming only under the
right group
\bea
& \tilde C_+=\tilde A_+,\,\,\,
\tilde C_-=-{\hat g}^{-1}\tilde A_-g+{\hat g}^{-1}\tilde D_- g, &\\
&\tilde C_\pm\longrightarrow {\hat R}^{-1}\tilde D_\pm  R+
{\hat R}^{-1}\tilde C_\pm  R. &
\eea
We now require the $\tilde B_\pm$ connexions to have zero curvature. This gives
ten  equations corresponding to the vanishing of the spinor-spinor components
of the curvature. Two of these equations simply determine the vector component
of the connexion in terms of the spinor components
\beq 2B_{++}=D_+\bar B_++\bar D_+ B_+-[\hat B_+,\bar B_+\},
\eeq
and there is an analogous equation determining $B_{--}$. These are just the
familiar conventional constraints of super-Yang-Mills theories. Among the eight
remaining equations, four involve only one light-cone chirality
\bea
D_+B_+-\hat B_+B_+ &=& 0\Leftrightarrow D_+A_++\hat A_+A_+=0,\label{eq110}\\
D_-B_--\hat B_-B_- &=& 0\Leftrightarrow D_-A_--\hat A_-A_-=0,
\label{eq10}\eea
and there are similar equations for $\bar B_+$, $\bar B_-$, or, equivalently,
for $\bar A_+$, $\bar A_-$. These again are well-known constraints of
super-Yang-Mills theories, usually referred to as representation preserving
constraints. Finally, there remains four equations involving both light-cone
chiralities
\beq
\tilde D_+\tilde B_-+\tilde D_-\tilde B_+-[\hat{\tilde B}_+,\tilde B_-\}=0.
\label{eq11}\eeq
In order to recover the Toda equations in this framework, one has to add to the
zero curvature equations some gauge invariant constraints. These constraints
involve the roots of the principal $sl(2)$ embedding in the bosonic case, and
the simple fermionic roots of the principal $osp(2\vert 1)$ embedding in the
$N=1$ supersymmetric case. Here it involves the simple fermionic roots of the
principal $sl(2\vert 1)$ embedding, e.g.
\beq
(\tilde B_+)_{>0}=\tilde\mu_+,\,\,\,
(\tilde C_-)_{<0}=\tilde\mu_-.
\label{eq12}\eeq
The gauge invariance of these constraints as usual follows from the fact that
$\mu_+$ and $\bar\mu_+$ ($\mu_-$ and $\bar\mu_-$) belong to the space
$\cg_{1\over 2}$ ($\cg_{-{1\over 2}}$) with smallest positive (negative) grade.
Before establishing the relation with the Toda equations as given in \cite{EH},
let us study the superconformal invariance of our equations. Beside gauge
invariance, let us note that the zero curvature equations possess two
additional symmetries. The first is a loop invariance, where the infinitesimal
parameters
$l_0$ and $r_0$ belong to the grade zero subalgebra $\cg_0$. It acts on the
fields by
\beq
\delta g=l_0g+gr_0,\,\,\delta\tilde A_+=[\tilde A_+, r_0],\,\,
\delta \tilde A_-=[l_0,\tilde A_-],\label{eq17}  \eeq
The parameters depend only on one light-cone chirality,
\beq  D_-l_0=\bar D_-l_0=0,\,\,\,D_+ r_0=\bar D_+ r_0=0.  \eeq
The second symmetry is \n2 superconformal invariance. Let us denote by
$K(\epsilon^{++},\epsilon^{--})$ the differential operator
\bea
K(\epsilon^{++},\epsilon^{--})&=&\epsilon^{++}\partial_{++}+
{1\over 2}D_+\epsilon^{++}\bar D_+
+{1\over 2}\bar D_+\epsilon^{++} D_+\nonumber\\
&+& \epsilon^{--}\partial_{--}+
{1\over 2}D_-\epsilon^{--}\bar D_-
+{1\over 2}\bar D_-\epsilon^{--} D_-
,\eea
where the infinitesimal parameters $\epsilon^{\pm\pm}$ depend only on one
light-cone chirality,
\beq D_-\epsilon^{++}=\bar D_-\epsilon^{++}=0,\,\,\,\,
D_+\epsilon^{--}=\bar D_+\epsilon^{--}=0.
\eeq
The superconformal transformations of the scalar superfield $g$ and of the
spinor components of the connexion are
\bea
\delta g &=& K(\epsilon^{++},\epsilon^{--})g,\nonumber \\
\delta A_\pm &=& K(\epsilon^{++},\epsilon^{--})A_\pm+{1\over 2}
D_\pm\bar D_\pm\epsilon^{\pm\pm}A_\pm,\label{eq30}\\
\delta\bar A_\pm &=& K(\epsilon^{++},\epsilon^{--})\bar A_\pm+{1\over 2}
\bar D_\pm D_\pm\epsilon^{\pm\pm}\bar A_\pm.
\nonumber\eea
The constraints \reff{eq12} are not invariant separately under Kac-Moody and
superconformal transformations. There is however a unique choice of the
Kac-Moody parameters $l_0$ and $r_0$ in terms of the superconformal
parameters $\epsilon^{++}$ and $\epsilon^{--}$ such that the combined
transformations leave the constraints invariant. It is given by
\beq
l_0=-D_+\bar D_+\epsilon^{++}\bar H-\bar D_+D_+\epsilon^{++} H,\,\,
r_0=-D_-\bar D_-\epsilon^{--}\bar H-\bar D_-D_-\epsilon^{--} H.
\label{eq18}\eeq
Thus we conclude that the constraints \reff{eq12} do allow for \n2
superconformal invariance, provided the superconformal transforation laws are
suitably modified.

\subsection{Relation with \n2 Toda equations}

This relation is most easily obtained by choosing the particular gauge
where the superfield $g$ is constrained to belong to the subgroup $G_0$,
\beq  g=g_0=\mbox{exp}(\sum_{i=1}^{2n}\phi^iH_i)\eeq
In this gauge, the constraints \reff{eq12} simply determine
the spinor connexions
\beq
\tilde A_+=-g_0^{-1}\tilde\mu_+ g_0,\,\,\, \tilde A_-=-g_0\tilde\mu_- g_0^{-1}.
\eeq
We then have to take into account the fact that these connexions are
constrained superfields. They satisfy eqs.(\ref{eq110},\ref{eq10}), which may
be translated into the following constraints on $g_0$
\bea [\mu_+,D_+g_0 {g_0}^{-1}]=0, & &
     [\bar\mu_+,\bar D_+g_0 {g_0}^{-1}]=0,\cr
     [ \mu_- ,g_0^{-1} D_- g_0 ] =0, & &
     [\bar\mu_-,{g_0}^{-1}\bar  D_-g_0 ]=0.    \eea
 More explicitly, the fields $\phi^i$ satisfy the following supersymmetric
chirality constraints
\beq \bar D_+\phi^{2i-1}=\bar D_-\phi^{2i-1}=0,\,\,
D_+\phi^{2i}= D_-\phi^{2i}=0.
\eeq
Among the four zero curvature equations \reff{eq11}, there remains only two
dynamical equations
\beq D_-(D_+g_0 {g_0}^{-1})+\{\mu_+,g_0\mu_- {g_0}^{-1}\}=0,\,\,\,
     \bar D_-(\bar D_+g_0 {g_0}^{-1})+\{\bar\mu_+,g_0\bar\mu_- {g_0}^{-1}\}=0,
\eeq
which, when expressed on the fields $\phi^i$, become the \n2 Toda equations
given in \cite{EH},
\beq
D_+D_-\phi^{2j-1}=(n-j+1)e^{\phi^{2j}-\phi^{2j-2}},\,\,\,
\bar D_+\bar D_-\phi^{2j}=-je^{\phi^{2j+1}-\phi^{2j-1}}.
\eeq
The next section will be devoted to a discussion of conserved currents and
of various Drinfeld-Sokolov type gauges.

\subsection{Conserved currents and gauge choices}

{}From the zero curvature equations it is clear that any gauge invariant
function $F(B_+,\bar B_+)$ or $G(C_-,\bar C_-)$ will satisfy the
light-cone chirality conditions
\beq
D_-F(B_+,\bar B_+)=\bar D_-F(B_+,\bar B_+)=0,\,\,\,
D_+G(C_-,\bar C_-)=\bar D_+G(C_-,\bar C_-)=0.
\eeq
Let us concentrate on the functions $F(B_+,\bar B_+)$. We shall now find a
generating set for these gauge invariant functions \cite{DS}. The elements of
this set are differential polynomials in the matrix elements of
 $(B_+)_{\leq 0}$,
$(\bar B_+)_{\leq 0}$. These polynomials will be found by exhibiting a unique
gauge transformation which brings the connexions to a prescribed form. In order
to do this, we use our knowledge of the $N=1$ case \cite{DRS}.
We first consider the sum
$B_++\bar B_+$, which is constrained by eq.\reff{eq12} to satisfy
$(B_++\bar B_+)_{>0}=f_+$. The adjoint action of the
$osp(2\vert 1)$ generator $f_+$ on the gauge algebra $\cg_-$ is non degenerate.
Let us separate $\cg_{\leq 0}$ into $(\mbox{\rm Im ad} f_+)_{\leq 0}$ plus some
graded supplementary subspace $V_{\leq 0}$. This space has one basis element
$T_{-{i\over 2}}$ at each strictly negative grade. Then there
exist a unique gauge transformation $e^F\in G_{<0}$ such that
\beq
e^{\hat F}(D_++\bar D_++(B_+)_{\leq 0}+(\bar B_+)_{\leq 0} +f_+)e^{-F}
=D_++\bar D_++W+f_+,\label{eq14}\eeq
and $W=\sum_{i=1}^{2n} W_{-{i\over 2}}T_{-{i\over 2}}$ belongs to $V_{\leq 0}$.
Moreover, both $F$ and $W$ are differential polynomials in the matrix elements
of $(B_+)_{\leq 0}$ and $(\bar B_+)_{\leq 0}$. It remains to be seen whether
all polynomials $W_{-{i\over 2}}$ are independent or not. To do this, we
restrict to the special case when the basis element of $V_{\leq 0}$ at
half-integer grade
$T_{-i+{1\over 2}}$ are related to the basis element at integer grade
by
\beq
T_{-i+{1\over 2}}=\alpha_i[\mu_+,T_{-i}]+\beta_i[\bar\mu_+,T_{-i}],\,\,\,
\alpha_i\neq\beta_i .\eeq
Moreover, we require the sets $\{T_{-i},[\mu_+,T_{-i}], \,1\leq i\leq n\}$
and $\{T_{-i},[\bar\mu_+,T_{-i}], \,1\leq i\leq n\}$ to span abelian
superalgebras. We shall exhibit later three choices for the space $V_{\leq 0}$
satisfying these requirements. Then it may be shown iteratively that
eq.\reff{eq14}, together with the non-linear constraints
(\ref{eq110},\ref{eq10}) completely
determine the gauge-fixed form of $(B_+)_{\leq 0}$ and
$(\bar B_+)_{\leq 0}$ to be
\bea (B_+)_{\leq 0}^{\mbox{\scriptsize gf}} &=&
\sum_{i=1}^{n}\alpha_i(-D_+W_{i+{1\over 2}}T_{-i}+W_{i+{1\over
2}}[\mu_+,T_{-i}]),\\
(\bar B_+)_{\leq 0}^{\mbox{\scriptsize gf}} &=& \sum_{i=1}^{n}\beta_i(-\bar
D_+W_{i+{1\over 2}}T_{-i}+W_{i+{1\over 2}}[\bar\mu_+,T_{-i}]).\eea
{}From this we conclude that there are only $n$ independent polynomials
$W_{i+{1\over 2}}$, $1\leq i\leq n$. This is half the number of conserved
currents found in the $N=1$ framework, which was to be expected since an
unconstrained $(2,0)$ superfield contains two $(1,0)$ superfields.
Let us give examples of gauges satisfying our requirements. The first was
given in \cite{EH}, it is a vertical gauge where $T_{-i}=E_{2i+1,1}$,
$T_{-i+{1\over2}}=E_{2i,1}=[\bar\mu_+, T_{-i}]$. Then the gauge-fixed forms are
\beq (B_+)_{\leq 0}^{\mbox{\scriptsize gf}}=0,\,\,
(\bar B_+)_{\leq 0}^{\mbox{\scriptsize gf}}=\sum_{i=1}^{n}(-\bar
D_+V_iE_{2i+1,1}
+V_i E_{2i,1}).\label{eq40}\eeq
One may construct as well a horizontal gauge by choosing
$T_{-n+i-1}=E_{2n+1,2i-1}$
and $T_{-n+i-{1\over2}}=E_{2n+1,2i}=-[\mu_+, T_{-n+i-1}]$. The gauge-fixed
forms are
\beq (B_+)_{\leq 0}^{\mbox{\scriptsize gf}}=\sum_{i=1}^{n}( D_+X_iE_{2n+1,2i-1}
+X_i E_{2n+1,2i}),\,\,
(\bar B_+)_{\leq 0}^{\mbox{\scriptsize gf}}=0.\eeq
A third possibility is to take an $osp(2\vert 1)$ lowest weight gauge, that is
to say $T_{-i}=(E_{--})^i$ and $T_{-i+{1\over2}}=[\mu_+-\bar\mu_+,(E_{--})^i]$,
where the matrix $E_{--}$ has been defined in eq.\reff{eq15}.
Then both gauge-fixed forms are non-zero and read
\bea (B_+)_{\leq 0}^{\mbox{\scriptsize gf}} &=&
\sum_{i=1}^{n}(D_+Z_iT_{-i}+Z_i[\mu_+,T_{-i}]),\\
(\bar B_+)_{\leq 0}^{\mbox{\scriptsize gf}} &=& -\sum_{i=1}^{n}(\bar
D_+Z_iT_{-i}+Z_i[\bar\mu_+,T_{-i}]).\label{eq16}\eea
It is in this gauge that the conserved currents have simple \n2 superconformal
transformations. From eqs.(\ref{eq17},\ref{eq30},\ref{eq18}) we find the
superconformal transformations of $B_+$ and $\bar B_+$ to be
\bea
\delta B_+ &=& K(\epsilon^{++},\epsilon^{--})B_++{1\over 2}D_+\bar
D_+\epsilon^{++}B_{+}+[l_0,B_+]+D_+l_0\\
\delta\bar B_+ &=& K(\epsilon^{++},\epsilon^{--})\bar B_++{1\over 2}\bar D_+
D_+\epsilon^{++}\bar B_{+}+[l_0,\bar B_+]+\bar D_+l_0
\eea
The lowest weight gauge \reff{eq16} is not stable under these transformations.
However, only a very simple compensating gauge transformation $1+\delta G$ is
needed,
where $\delta G$ is non zero only at grades $-{1\over 2}$ and $-1$. One finds
that all currents beside $Z_1$ are \n2 superprimary fields,
\beq
\delta Z_j=K(\epsilon^{++},\epsilon^{--})Z_j +j\partial_{++}\epsilon^{++}Z_j,
\eeq
and $Z_1$ has the transformation law of a super-energy-momentum tensor
\beq
\delta Z_1=K(\epsilon^{++},\epsilon^{--})Z_1+\partial_{++}\epsilon^{++}Z_1+
{1\over 2}[D_+,\bar D_+]\partial_{++}\epsilon^{++}.
\eeq
The \n2 Miura transformation, expressing the conserved currents $V_i$ in
eq.\reff{eq40} in terms of the Toda superfield $\phi^i$ is discussed in
\cite{EH}. We will meet this transformation again in the context of the \n2
supersymmetric KdV equation in  section $3$.

\subsection{Non-abelian \n2  Toda equations}

It is clear how the methods developped in \cite{EH}
and in the present article can be generalized to the case of non-abelian Toda
equations. We shall not try to construct a general theory here, but restrict
our attention to the Toda equations associated with an $sl(2\vert 1)$ embedding
inside $sl(n+1\vert n)$. Such equations have the property that one can find a
generating set for the conserved currents such that all elements beside the
\n2 superconformal tensor are \n2 super-primary fields. The classification of
$sl(2\vert 1)$ embeddings inside a simple superalgebra has been established in
\cite{RSS}. In our case, it amounts to decompose the $2n+1\times 2n+1$ matrices
into blocks  so that the size of the diagonal blocks is odd. These diagonal
blocks form a regular subalgebra. The
semi-simple part of this subalgebra is a sum of superalgebras of the type
$sl(n_i+1\vert n_i)$ or $sl(m_i\vert m_i+1)$. We then consider the
$sl(2\vert 1)$ embedding which is principal in this superalgebra. Thus $\mu_+$
($\bar\mu_+$) is still the sum of those matrices $E_{2i-1,2i}$ ($E_{2i,2i+1}$)
which are inside the diagonal blocks. All this is very reminiscent of the
construction of $sl(2)$ subalgebras of $sl(n)$. A difference is that in the
case of $sl(2\vert 1)$ embeddings, the ordering of the blocks is of some
importance. For examples, the decompositions $5=3+1+1$ and $5=1+3+1$ lead to
different embeddings. Just as in the principal case, the $osp(1\vert 2)$
generator $e_0=H+\bar H$ induces a half-integer gradation of the superalgebra
$sl(n+1\vert n)$, and $\cg_0$ is the subalgebra of grade zero. The Toda
superfield $g_0(x,\theta,\bar\theta)$ belongs to the corresponding supergroup
$G_0$. We may then consider the connexions
\beq \tilde B_+=\tilde D_+g_0g_0^{-1}+\tilde\mu_+,\,\,\,
     \tilde B_-=-\hat g_0\tilde\mu_- g_0^{-1}. \eeq
Then we write for these connexions the zero curvature equations
(\ref{eq110},\ref{eq10},\ref{eq11}). The equations involving only
one light-cone chirality lead to the constraints
\bea [\mu_+, D_+g_0g_0^{-1}\} =0,&& [\bar\mu_+,\bar D_+g_0g_0^{-1}\}
=0,\cr
  [\mu_-,\hat g_0^{-1} D_-g_0\} =0, && [\bar\mu_-,\hat g_0^{-1}\bar
D_-g_0\} =0. \label{eq61} \eea
The equations \reff{eq12} involving both light-cone chiralities lead to the
 four dynamical equations
\beq\tilde D_-(\tilde D_+g_0g_0^{-1})+\tilde\mu_+\hat g_0\tilde\mu_-g_0^{-1}
+g_0\tilde\mu_-\hat g_0^{-1}\tilde\mu_+=0.  \eeq
This is not the end of the story, since one can easily check that, even
taking into account the constraints \reff{eq61}, the
superfield $g_0$ contains
too many components. The situation here is the same as in
 the \n2 WZNW model \cite{HS},
and we expect the constraints on the super-currents to be constructed from two
classical r-matrices $r_L$ and $r_R$ of the superalgebra $\cg_0$. An $r$-matrix
$r=r_L$ or $r=r_R$ is a linear transformation of $\cg_0$
satisfying the classical modified Yang-Baxter equation
\beq r[rM,N\}+r[M,rN\}-[rM,rN\}=[M,N\}.\eeq
Moreover, we require $r$ to be a super-antisymmetric involution
\beq r^2=Id,\,\,\, \tr(rM)N=-\tr M(rN).\eeq
We  note  $\cg_0^+$ and $\cg_0^-$ the eigenspaces of $r$ with respective
eigenvalues $1$ and $-1$. Then the properties of $r$ show that $\cg_0^+$ and
$\cg_0^-$ are isotropic subalgebras of $\cg_0$. The data ($\cg_0$, $\cg_0^+$,
$\cg_0^-$) is called a Manin triple.

Let us call
$\cg^+_L$ and $\cg^+_R$ the eigenspaces of $r_L$ and $r_R$ with eigenvalue $1$,
$\cg^-_L$ and $\cg^-_R$ the eigenspaces of $r_L$ and $r_R$ with eigenvalue
$-1$.
Then we expect the form of the constraints on the super-currents to be
\bea D_+g_0g_0^{-1}\in\cg^+_R, & \,\,\, & \bar D_+g_0g_0^{-1}\in\cg^-_R,
\nonumber\\
\hat g_0^{-1} D_-g_0\in\cg^+_L, & \,\,\, &\hat g_0^{-1}\bar D_-g_0\in\cg^-_L,
\label{eq62}\eea
The possible choices for the r-matrices are severely restricted if we require
that the constraints \reff{eq61}  coming from the zero curvature equations
should be a subset of the complete constraints  \reff{eq62}. This turns
out to be easily realized. We use the decompositions of the superalgebra
$\cg_0$ as
\beq \cg_0= (\mbox{Im ad}\mu_+)_0\oplus(\mbox{Im
ad}\bar\mu_+)_0\oplus\cg_0^S = (\mbox{Im ad}\mu_-)_0\oplus(\mbox{Im
ad}\bar\mu_-)_0\oplus\cg_0^S, \eeq
where $\cg_0^S$ is the singlet part of $\cg_0$ under the $sl(2\vert 1)$ action.
The set $\mbox{Im ad}\mu_+$ is easily shown to be a superalgebra on which the
invariant quadratic form vanishes. Then we may take
\beq (\mbox{Im ad}\mu_+)_0\subset\cg^+_R,\,\, \mbox{(Im
ad}\bar\mu_+)_0\subset\cg^-_R,\,\,
(\mbox{Im ad}\mu_-)_0\subset\cg^+_L,\,\, (\mbox{Im
ad}\bar\mu_-)_0\subset\cg^-_L. \eeq
The subalgebra $\cg_0^S$ is orthogonal to the spaces $\mbox{Im ad}\mu_\pm$,
$\mbox{Im ad}\bar\mu_\pm$, and one has the commutation relations
\beq
[\cg_0^S,\mbox{Im ad}\mu_\pm\}\in \mbox{Im ad}\mu_\pm,\,\,
[\cg_0^S,\mbox{Im ad}\bar\mu_\pm\}\in \mbox{Im ad}\bar\mu_\pm.
\eeq
{}From these equations we conclude that we may choose freely the definition of
$r_R$  and $r_L$ inside $\cg_0^S$. We may for instance take the upper (lower)
triangular elements in $\cg_0^S$ to be included in $\cg^+_{R,L}$
($\cg^-_{R,L}$). We denote by $\mbox{\bf 1}_i$ the identity matrix in the $i$th
block. Then the matrices $H_i=\mbox{\bf 1}_i+\mbox{\bf 1}_{i+1}$ span the even
dimensional Cartan
subalgebra of $\cg_0^S$. We may take $H_{2i-1}\in\cg^+_{R,L}$ and
$H_{2i}\in\cg^-_{R,L}$. With these choices, the complete constraints
\reff{eq62} imply the constraints \reff{eq61}.

The construction of a gauge invariant formulation of the \n2 non-abelian Toda
equations follows exactly the same line as the abelian case. The only
difference is that beside the constraints \reff{eq12}, one should add gauge
invariant constraints on the $sl(2\vert 1)$ singlet part of $B_+$, $\bar B_+$,
$C_-$, $\bar C_-$, of the form
\beq
 (B_+)_S\in\cg^+_R,  \,\,\,  (\bar B_+)_S\in\cg^-_R
,\,\,\, (C_-)_S\in\cg^+_L, \,\,\,  (\bar C_-)_S\in\cg^+_L.
\eeq

\section{\n2 KdV equation in superspace}
\setcounter{equation}{0}

The methods used in the first section for the description of the \n2
supersymmetric Toda equations in extended superspace apply as well to the \n2
KdV equation. The formulation in \n2 superspace that we shall give
is strongly inspired from the $N=1$ treatment given in \cite{IK}. Most of the
notations that we use also come from \cite{IK}.
 \subsection{Lax operators, gauge invariance}
The relevant superalgebra is now the untwisted loop algebra constructed from
$sl(2\vert 2)$,
or rather the quotient $A^{(1)}(1\vert 1)$ of this algebra by its center. We
thus consider the set of the $4\times 4$ real matrices depending on a loop
parameter $\lambda$. As in the first section, the supertrace is defined as the
alternate sum of diagonal elements, and we consider the superalgebra of
matrices with zero supertrace. Any function of $\lambda$ multiplying the
identity matrix is in the center of this algebra.  We choose a representative
in each equivalence class of the quotient by the center by requiring the last
element on the diagonal to vanish. We introduce the following odd elements
\beq
\omega=\left(
\begin{array}{cccc} 0 & 1 & 0 & 0\\ 0 & 0 & 0 & 0\\ 0 & 0 & 0 & 1\\
0 & 0 & 0 & 0\\
\end{array}\right),\,\,
\bar\omega=\left(
\begin{array}{cccc} 0 & 0 & 0 & 0\\ 0 & 0 & 1 & 0\\ 0 & 0 & 0 & 0\\
\lambda & 0 & 0 & 0\\
\end{array}\right),\,\,
\Lambda=\omega+\bar\omega,
\eeq
which satisfy
\beq \omega^2=0,\,\,\,\bar\omega^2=0,\,\,\,\{\omega,\bar\omega\}=\Lambda^2.
\eeq
$\Lambda^2$ is an even semi-simple element of the superalgebra. We shall use
the notations
\beq {\ck}=\mbox{ker ad}\Lambda^2,\,\,
{\ck}=[\ck,\ck\}\oplus\ck',
\eeq
where $[\ck,\ck\}$ denotes the commutator subalgebra
 of $\ck$, and $\ck'$ some complement.
We consider a gradation $d$ of the superalgebra defined by
$d(\lambda^pE_{i,j})=4p+j-i$. Notice that $\omega$ and $\bar\omega$ have grade
$1$, and $\Lambda^2$ has grade $2$.
The \n2 superspace has a bosonic coordinate $x$ and two
Grassmann coordinates $\theta$, $\bar\theta$. Supersymmetric covariant
derivatives are
\beq D={\partial\over\partial\theta}+\bar\theta\partial,\,\,
\bar D={\partial\over\partial\bar\theta}+\theta\partial,\,\,
\partial={\partial\over\partial x}.\label{65}\eeq
In analogy with the methods developped in
\cite{DS}, we introduce the spinor Lax operators \beq \cl=D+q+\omega,\,\,\,
\bar\cl=\bar D+\bar q+\bar\omega,\label{eq63}\eeq
where the superfields $q$ and $\bar q$ are $\lambda$-independent odd
matrices of
non-positive grade. In particular, the upper triangular elements of $q$ and
$\bar q$ vanish. The operators $\cl$ and $\bar\cl$ are required to
satisfy zero curvature
equations. Two of these equations are constraints on $q$ and $\bar q$
\bea
\hat\cl\cl = 0 &\Rightarrow& Dq+\hat qq+[\hat q,\omega\}=0,\label{eq164}\\
\hat{\bar\cl}\bar\cl = 0 &\Rightarrow&\bar D\bar q+\hat{\bar q}\bar q+[\hat
{\bar q},\bar\omega\}=0,\label{eq64}\eea
while the third simply determines the vector Lax operator
\beq \cl_x=\hat\cl\bar\cl+\hat{\bar\cl}\cl=2\partial+Q-\Lambda^2.\eeq
As in the Toda case, equations (\ref{eq164},\ref{eq64})
 may be interpreted as follows.
$\mbox{ad}\omega$ being a nilpotent element of the superalgebra, its image is
included in its Kernel. The elements in $q$ which do not belong to the Kernel
are fully determined in terms of those in the Kernel. Moreover, the elements in
$q$ belonging to the image of $\mbox{ad}\omega$ are  unconstrained \n2
superfields, while those belonging to the Kernel but not to the image satisfy
chirality-type constraints. Similar statements hold for $\bar q$ and
$\bar\omega$.

The form \reff{eq63} of the Lax operators and the zero curvature equations
(\ref{eq164},\ref{eq64}) are invariant under the gauge transformations
\beq
\cl\rightarrow\hat g\cl g^{-1},\,\,\, \bar\cl\rightarrow\hat g\bar\cl g^{-1},
\label{eq65}\eeq
where $g(x,\theta,\bar\theta)$ is a $\lambda$-independent lower triangular
matrix with $1$'s along the diagonal. Two particular gauge-fixings will be
useful in the sequel. One is a vertical gauge
\beq q^1_{gf}=0,\,\,\,\bar q^1_{gf}=\left(\begin{array}{cccc}
 0 & 0 & 0 & 0 \\  \bar DW & 0 & 0 & 0 \\  W & 0 & 0 & 0 \\
 0 & 0 & 0 & 0 \end{array}\right),\label{eq66}\eeq
where $W$ is an unconstrained \n2 superfield. As in the Toda case, the
existence of this gauge is shown by using the $N=1$ results established in
\cite{IK} and taking into account the constraints (\ref{eq164},\ref{eq64}).
Notice that this
gauge is not of Drinfeld-Sokolov type. That is to  say that the gauge group
element which brings $q$ and $\bar q$ to this form is not a local differential
polynomial in the matrix elements of $q$ and $\bar q$. The other gauge we shall
consider is a diagonal gauge
\beq q^2_{gf}=\left(\begin{array}{cccc}  \phi & 0 & 0 & 0 \\  0 & \phi & 0 & 0
\\
 0 & 0 & 0 & 0 \\  0 & 0 & 0 & 0 \end{array}\right),\,\,\,
\bar q^2_{gf}=\left(\begin{array}{cccc} 0 & 0 & 0 & 0 \\  0 & \bar\phi & 0 & 0
\\
 0 & 0 & \bar\phi & 0 \\ 0 & 0 & 0 & 0 \end{array}\right),\label{eq200}\eeq
where $\phi$ is chiral, $D\phi=0$, and $\bar\phi$ is anti-chiral,
$\bar D\bar\phi=0$. Although the gauge \reff{eq66} is not of Drinfeld-Sokolov
type, the superfield $W$ may be expressed as a differential polynomial in terms
of the superfields $\phi$ and $\bar\phi$. This is the celebrated Miura
transformation, which is most easily obtained by introducing a $4$-vector
$\psi=(\psi_1,\psi_2,\psi_3,\psi_4)^t$ which is annihilated by the Lax
operators
\beq\cl\psi=0,\,\,\,\bar\cl\psi=0.\label{eq67}\eeq
The first component $\psi_1$ of the vector $\psi$ is gauge invariant. The
matrix equations \reff{eq67} reduce to scalar gauge-invariant differential
equations on $\psi_1$. In the vertical gauge, one obtains
\beq \bar D\psi_1=0,\,\,\,\bar DD(\partial \psi_1-W\psi_1)=\lambda\psi_1,\eeq
and in the diagonal gauge
\beq \bar D\psi_1=0,\,\,\,\bar DD(\bar D+\bar\phi)(D+\phi)\psi_1
=\lambda\psi_1,\eeq
which leads to the relation $W=-(\phi\bar\phi+\bar D\phi+D\bar\phi)$.

\subsection{Evolution equations, conserved charges}
We wish to write evolution equations for the Lax operators $\cl$ and
$\bar\cl$ of the type
\beq{\partial\cl\over\partial t}=A\cl-\cl\hat A,\,\,
{\partial\bar\cl\over\partial t}=A\bar\cl-\bar\cl\hat A,\label{eq69}\eeq
and $A$ is an even matrix belonging to the superalgebra $A^{(1)}(1\vert 1)$.
As in the bosonic case \cite{DS}, we first need to construct an even
matrix $M$ commuting with $\cl$ and $\bar\cl$
\beq M\cl-\cl\hat M=0,\,\,\, M\bar\cl-\bar\cl\hat M=0 .\eeq
We decompose  $M$ as $M=M^++M^-$, where the grades in $M^+$ are positive
or zero, and the grades in $M^-$ are strictly negative. Then we can take
$A=M^+$. In order to construct the matrix $M$, we shall show that there exists
a
matrix $F=\sum_{n=1}^\infty F_{-n}$, $d(F_{-n})=-n$, which brings $\cl$ and
$\bar\cl$ to the form
\beq e^{\hat F}\cl e^{-F}=D+H+\omega,\,\,\,
e^{\hat F}\bar\cl e^{-F}=\bar D+\bar H+\bar\omega,  \eeq
where $H=\sum_{n=1}^\infty H_{-n}$ and $\bar H=\sum_{n=1}^\infty\bar H_{-n}$
belong to the Kernel $\ck$ of
$\mbox{ad}\Lambda^2$. At any finite grade, $F$, $H$ and $\bar H$ depend
polynomially on the matrix elements of $q$ and $\bar q$, and on their
derivatives. Suppose that the proof has been carried out down to the grade
$-n+1$
for $H$ and $\bar H$, and down to the grade $-n$ for $F$. At the next grade, it
will be convenient to look first at the sum $H+\bar H$. We have
\beq H_{-n}+\bar H_{-n}=P(H_{-p},\bar H_{-p}, F_{-p})+[\hat F_{-n-1},\Lambda\},
\eeq
where $P$ is a differential polynomial in $H_{-p}$, $\bar H_{-p}$ with $p<n$
and in $F_{-p}$ with $p\leq n$. We choose the particular form $ F_{-n-1}=\hat
G_{-n-2}\Lambda-\Lambda G_{-n-2}$. Then the last equation becomes
\beq H_{-n}+\bar H_{-n}=P(H_{-p},\bar H_{-p}, F_{-p})+[\Lambda^2,
G_{-n-2}]. \eeq
We then use the fact that $\Lambda^2$ is semi-simple to conclude that we may
choose $G_{-n-2}$ in such a way that $H_{-n}+\bar H_{-n}$ belongs to $\ck$. To
show that both $H_{-n}$ and $\bar H_{-n}$ belong to $\ck$, we use the
constraints (\ref{eq164},\ref{eq64}) coming from the zero curvature equations.
They lead at
grade $-n+1$ to the equations
\bea
& D H_{-n+1}+\sum_{p+q=n-1}[\hat H_{-p},H_{-q}\}
+[\hat H_{-n},\omega\} = 0 &\label{eq167} \\
& \bar D\bar  H_{-n+1}+\sum_{p+q=n-1}[\hat{\bar  H}_{-p},\bar H_{-q}\}
+[\hat{\bar H}_{-n},\bar\omega\} = 0, &
\label{eq68}\\
& \Rightarrow
[\hat H_{-n},\omega\}\in {\ck},\,\,
[\hat{\bar H}_{-n},\bar\omega\}\in\ck. &\eea
Thus we know that $H_{-n}+\bar H_{-n}$, $[\hat H_{-n},\omega\}$ and
$[\hat{\bar H}_{-n},\bar\omega\}$ are in $\ck$. Taking a supercommutator of
$H_{-n}+\bar H_{-n}$ with $\omega$ or $\bar\omega$ shows that also
$[\hat H_{-n},\bar\omega\}$ and $[\hat{\bar H}_{-n},\omega\}$ are in
$\ck$. Then, using the fact that $\Lambda^2=\{\omega,\bar\omega\}$ is
semi-simple, we conclude that
 $H_{-n}$ and $\bar H_{-n}$ belong to
$\ck$. It will be useful to split $H$ and $\bar H$ as
\beq H=C+H',\,\, \bar H=\bar C+\bar H',\,\, C,\bar C\in [{\ck},{\ck}\},\,\,
H',\bar H'\in\ck'\eeq
On $H'$ and $\bar H'$, the constraints (\ref{eq167},\ref{eq68}) reduce to the
chirality conditions $DH'=0$, $\bar D\bar H'=0$. Notice also that $e^F$ is only
defined up to a left multiplication by $e^T$, with $T$ in $\ck$. As a
consequence, $H'$ and $\bar H'$ are only defined up to the addition of a total
derivative
\beq H'\rightarrow H'+DT',\,\,\, \bar H'\rightarrow \bar H'+\bar DT'.\eeq
Then, the integrated quantities
\bea Q=\int dV_c H' &=& \int dx(\bar DH')_{\theta=\bar\theta=0}\\
\bar Q=\int dV_a H' &=& \int dx(D\bar H')_{\theta=\bar\theta=0} \eea
are not uniquely defined
\beq Q\rightarrow Q+\int dx(\bar DDT')_{\theta=\bar\theta=0},\,\,\,
\bar Q\rightarrow \bar Q+\int dx(D\bar DT')_{\theta=\bar\theta=0}.\eeq
However, the sum $Q+\bar Q$ is uniquely defined
\beq Q+\bar Q\rightarrow Q+\bar Q+\int dx(2\partial T')_{\theta=\bar\theta=0}
=Q+\bar Q.\eeq
The fact that this quantity is uniquely defined also implies that it is a gauge
invariant functional of $q$ and $\bar q$. As we shall see next, $Q_{-n}+\bar
Q_{-n}$ are the conserved charges of the \n2 KdV hierarchy. From now on, things
work as in the bosonic case. We introduce the matrix
\beq M=e^{-\hat F}be^{\hat F},\eeq
where $b$ is a constant matrix in the center of $\ck$.
Then we choose
\beq A=M^+=(e^{-\hat F}be^{\hat F})^+.\eeq
The equations \reff{eq69} should be seen as evolution equations for gauge
invariant differential polynomials of $q$ and $\bar q$. Let us study these
equations on $H$ and $H'$. Using the notation
\beq B=e^{\hat F}{\partial\over \partial t}e^{-\hat F}+e^{\hat F}Ae^{-\hat F},
\eeq
we obtain
\bea
{\partial H\over \partial t}+DB &=& B(H+\omega)-(H+\omega)\hat B\\
{\partial\bar H\over \partial t}+\bar DB &=& B(\bar H+\bar \omega)
-(\bar H+\bar \omega)\hat B\eea
It is a consequence of these equations that $B$ belongs to $\ck$. Then the
right-hand side of these equations belongs to $[\ck,\ck\}$, and if we restrict
to $\ck'$ we find
\beq {\partial H'\over \partial t}+DB'=0,\,\,\,
{\partial\bar H'\over \partial t}+\bar DB'=0,\eeq
which implies that $Q+\bar Q$ is time independent.

We studied in some detail the case $b=\lambda\Lambda^2$, which has grade $6$.
We computed the evolution equation for the superfield $W$ appearing in the
gauge choice \reff{eq66}. When computed on some gauge-fixed form of the Lax
operators,
the evolution equations \reff{eq69} acquire an extra term
\beq{\partial\cl_{gf}\over\partial t}=(A+R)\cl_{gf}-\cl_{gf}(\hat A+\hat
R),\,\,
{\partial\bar\cl_{gf}\over\partial t}=(A+R)\bar\cl_{gf}-\bar\cl_{gf}(\hat
A+\hat R),\label{eq169}\eeq
where $R$ belongs to the gauge algebra. The computation of the objects $A$, $F$
and $R$ is somewhat lengthy. The final result is
\beq {\partial W\over \partial t}=2\partial^3W-{3\over 2}\partial(DW\bar
DW)-{1\over4}\partial W^3\eeq
This \n2 KdV equation was obtained in \cite{Mat1}. Extensions of these results
to the KdV hierarchies associated with the superalgebras $sl(n\vert n)$ is in
principle straightforward.

\section*{Conclusion}

Let us consider the relation between our gauge invariant equations  and the
WZNW models. In the bosonic \cite{FOR} or $N=1$ \cite{DRS} cases, the gauge
invariant field equations may be considered as the field equations of a gauged
WZNW model. This means in particular that if the gauge fields are set to zero,
and the field equations
coming from the variation of these gauge fields are dropped, one recovers the
equation of motion of the WZNW model. In our case, since we do not have an
action, it is not clear which are the equations coming from the variation of
the gauge superfields. Naively, and in analogy with the bosonic case, one could
expect that the field equations of the connexions are the constraints
\reff{eq12}. If these constraints are dropped, and the gauge fields are set to
zero, one ends up with the following equations
\beq\tilde D_-(\tilde D_+ gg^{-1})=0.
\label{eq1000}\eeq
These are very nice equations, possessing a very big loop invariance, but they
are not the \n2 WZNW equations. In particular, the \n2 unconstrained superfield
$g$ contains a dynamical vector field. The true \n2 WZNW equations \cite{HS}
contain equation \reff{eq1000}, together with constraints on the superfield $g$
constructed from two classical $r$-matrices. These constraints cannot be
considered as a subset of the gauge-invariant constraints \reff{eq12}.
Thus at present the relation of our gauge invariant formulation with a
gauged WZNW model is unclear.

It is not a very big surprise to verify that methods which work in ordinary
space and in $N=1$ superspace do extend to \n2 superspace at the level of field
equations. It is considerably more difficult to obtain in extended superspace
actions or Hamiltonian formulations. However, as already stressed, an
Hamiltonian reduction approach to \n2 $\cal W$ algebras has been developped in
\cite{ivan2}.
This approach makes use of the \n2 operator product expansions for constrained
supercurrents constructed in \cite{HS}. These operator product expansions, or
rather the Poisson brackets which may be derived from them, would be relevant
in a Hamiltonian formulation of the non-abelian \n2 Toda equations discussed at
the end of section 2. It is however not clear to us how they could be used in
the
gauge-invariant approach to Toda or KdV equations.

\vspace*{1.0cm}

{{\bf Acknowledgements}\\
We thank E. Ivanov and P. Sorba for useful discussions.}

\end{document}